# 'Spillout' effect in gold nanoclusters embedded in c-$Al_2O_3$(0001) matrix

S. Dhara,[*] B. Sundaravel, T. R. Ravindran, K. G. M. Nair, C. David, B. K. Panigrahi,

and P. Magudapathy

Materials Science Division, Indira Gandhi Centre for Atomic Research, Kalpakkam –603 102, India

K. H. Chen

Institute of Atomic and Molecular Sciences, Academia Sinica, Taipei-106, Taiwan

Gold nanoclusters are grown by 1.8 MeV $Au^{++}$ implantation on c-$Al_2O_3$(0001) substrate and subsequent air annealing at temperatures ≥ 1273K. Post-annealed samples show plasmon resonance in the optical (579-561 nm) region for average cluster radius ~ 0.86-1.2 nm. A redshift of the plasmon peak with decreasing cluster size in the post-annealed samples is assigned to the 'spillout' effect (reduction of electron density) for clusters with ~157-427 number of Au atoms fully embedded in crystalline dielectric matrix with increased polarizability in the embedded system.

---

Electronic mail : dhara@igcar.ernet.in

Optical properties of noble metal nanoclusters embedded in dielectric matrices have attracted considerable attention in recent years mainly owing to their applications in nonlinear optics [1], and understanding fundamental issues related to the electronic properties in the small clusters [2]. The main feature in the optical response is the surface plasmon excitation (collective oscillation of the conduction electrons), resulting in a resonance band in the absorption spectra. For gold as well as for the other noble metals, the surface plasmon resonance (SPR) occurs in the near-UV-Visible region. It is therefore interesting to use them for optical-device applications. Reduction of the electron density ('spillout' effect) and the screening of the interaction of valence ($s$) electrons by core ($d$)-electrons in noble metals shift the frequency of light absorbed either to the red or blue region, respectively, with decreasing size. A large number of experimental and theoretical results were reported for the red or blue shift of the SPR peak position with decreasing noble metal cluster size as an effect of embedding matrix and surrounding porosity [2-5]. Here, we will focus the results of Au nanoclusters embedded in $Al_2O_3$ (alumina) matrix. In fact, Au nanoclusters in alumina matrix system seems to be interesting as it shows prominent nonlinear response (~ $4 \times 10^{-5}$ esu) with respect to that in silica matrix (~ $2.5 \times 10^{-6}$ esu) [1,6]. There are only very few results on Au nanocluster formation in porous alumina matrix. So far, an experimental observation of blueshift of SPR peak position is only reported for Au (2-4 nm) clusters embedded in porous alumina matrix grown by co-deposition technique using pulsed laser ablation [4]. Effect of porosity surrounding Au clusters is taken into consideration in the TDLDA calculation to support the experimental observation [2,5]. For embedded Au clusters in $Al_2O_3$ a redshift, originating from the 'spillout' effect with increasing polarizability in the system, is predicted by time dependent local density approximation (TDLDA) calculation [5]. So far, there is no experimental evidences in support to the prediction.

Various techniques were employed to grow the embedded noble metal clusters and study their optical properties [4]. However, the formation of nanoclusters by ion beam irradiation is of recent interest on account of certain advantages it offers over other techniques, which include excellent control of size and selectivity of depth, and area in which the nanoclusters form [7,8]. In the present study, we report the growth of Au nanoclusters in c-$Al_2O_3$ matrix using 1.8 MeV $Au^{++}$ implantation with low current density at various fluence ranges and subsequent annealing at high temperatures. The plasmon resonance spectra were studied using UV-Vis absorption studies. Cluster sizes were determined using acoustic phonon confinement studies by the low frequency Raman technique. A redshift of the SPR peak position with decreasing cluster size is reported in the small cluster sizes (radii ~ 0.86-1.2 nm) embedded in c-$Al_2O_3$ matrix. Our results of fully embedded Au nanoclusters in c-$Al_2O_3$ matrix differ completely from other studies where clusters were grown in porous alumina matrices [4,6].

Au clusters were grown by direct Au-ion implantation on c-$Al_2O_3$(0001) substrates. Implantation was performed using 1.8 MeV $Au^{++}$ at $1\times10^{-5}$ Pa with a low current density of ~12 mA $m^{-2}$ in the ion fluence range of $1\times10^{20}$-$1\times10^{21}$ $m^{-2}$. A low current density affects average size and size distribution when implanting Ag and Cu in dielectric matrix [9] and is expected from simulations [8] to play a similar role for gold. A 1.7 MV Tandetron accelerator (High Voltage Engineering Europa, The Netherlands) was used for the implantation study. High temperature air annealing at 1273K and 1473K for 1 hour was performed for studying the growth of the clusters. Optical absorption spectra were recorded at room temperature in the range of 200-1100 nm using a Hewlett-Packard diode-array UV-Vis spectrophotometer (Model 8453) with necessary signal correction for the substrate. Cluster sizes were determined using low-frequency Raman scattering studies at room temperature in the back scattering geometry, using vertically polarized 488 nm line of an argon ion laser (Coherent, USA) with 200 mW power. Two different configurations were employed, with the excitation and detection polarization either parallel (polarized) or perpendicular (depolarized) to each other. Polarized (*VV*) or depolarized (*VH*) scattered light from the sample was dispersed using a double monochromator (Spex, model 14018) with instrument resolution of 1.4 cm$^{-1}$ and detected using a cooled photomultiplier tube (FW ITT 130) operated in the photon counting mode.

The SPR frequencies are located in the UV-Visible (UV-Vis) spectral range at $\omega_{SPR} = \omega_P/[2\epsilon_m + \epsilon_d(\omega_{SPR})]^{1/2}$ (dipolar Mie resonance of a spherical metal cluster). $\omega_P = (\eta e^2/\epsilon_0 m_e^*)^{1/2}$ [$\eta$ the electron density, $\epsilon_0$ the dielectric function of bulk metal; $m_e^*$ the effective mass of electron] is the Drude free-electron plasma frequency. $\epsilon_d(\omega) = (1 + \chi^d)$ is the core-electron contribution [$\chi^d$ the interband part of dielectric susceptibility (*d* electrons)] to the complex dielectric function of the noble metal $\epsilon(\omega) = 1 + \chi^s(\omega) + \chi^d(\omega)$ [$\chi^s$ the Drude-Sommerfeld part of dielectric susceptibility (*s* electrons)]. $\epsilon_m$ ($\approx 3.1$ for c-Al$_2$O$_3$ in the relevant energy range) [10] is the dielectric function of matrix. A redshift is observed for very small clusters, as average electron density is reduced due to an increasing electron 'spillout' effect with decreasing cluster size. The physical idea underlying the blueshift trend observed in noble metal clusters is based on the assumption that due to the localized character of the core-electron wavefunctions the screening effects are less effective over a surface layer inside the metallic particle. Close to the surface the valence electrons are then incompletely embedded inside the ionic-core background. In two-region dielectric model, this hypothesis is taken into account by assuming that the effective polarizable continuous medium responsible for the screening does not extend over the whole cluster volume, by prescribing that $\chi^d(\omega) = \epsilon_d(\omega) - 1$ vanishes for radius $> R\text{-}r$; where '$R$' is the cluster radius and '$r$' a thickness parameter of the order of a fraction of the nearest-neighbor atomic distance.

The Mie frequency in the large-size limit for embedded Au nanoclusters in c-$Al_2O_3$ is approximately calculated by solving $\varepsilon(\omega) + 2\varepsilon_m(\omega) = 0$ and found to be ~ 2.3 eV (539 nm). Plasmon peaks around ~ 2.14-2.21 eV (~579-561 nm) are observed in the UV-Vis absorption study of the post-annealed samples for two different annealing conditions (Fig. 1) indicating formation of Au nanoclusters. Broadening of SPR peaks for the samples grown at lower fluences and annealed at lower temperature (1273K) is not understood fully and may be due to imperfections in the small crystallites and surrounding matrix. The peaks are quite sharp for the sample annealed at higher temperature (1473K) as Landau damping effect is expected to be small for fully embedded clusters [5]. For both the annealing conditions we observed a clear redshift of the SPR peak position with decreasing fluence. Incidentally, we could not observe any SPR peak in the as-grown samples. We start observing SPR peaks for the annealed samples where cluster may have grown into sufficiently bigger size for observable resonance to occur.

Low-frequency Raman study was performed for the determination of size and shape of clusters in the post-annealed samples. Confined surface acoustic phonons in metallic or semiconductor nanoclusters give rise to low-frequency modes in the vibrational spectra of the materials. The Raman active spheroidal motions are associated with dilation and strongly depend on the cluster material through $v_t$, the transverse and $v_l$, longitudinal sound velocities. These modes are characterized by two indices $l$ and $n$, where $l$ is the angular momentum quantum number and $n$ is the branch number. $n = 0$ represents the surface modes. The surface quadrupolar mode ($l = 2$, eigenfrequency $\eta^s_2$) appears in both the $VV$ and the $VH$ Raman scattering geometries whereas, the surface spherical mode ($l = 0$, eigenfrequency $\xi^s_0$) appears only in the polarized geometry.

Eigenfrequencies for the spheroidal modes at surface ($n = 0$; $l = 0, 2$) of Au nanocluster in $Al_2O_3$ matrix is calculated to be $\eta^s_2 = 0.84$ and $\xi^s_0 = 0.40$ by considering the matrix effect in the limit of elastic body approximation of small cluster (core-shell model) [11]. The surface quadrupolar mode frequencies corresponding to $l = 0$, and 2 are given by,

$$\omega^s_0 = \xi^s_0 v_l / Rc \,;\, \omega^s_2 = \eta^s_2 v_t / Rc \quad \ldots\ldots\ldots\ldots(1)$$

where $c$ is the velocity of light in vacuum, and $v_l = 3240$ m/sec, $v_t = 1200$ m/sec in Au. Figure 2 shows the Raman spectra of post-annealed samples prepared at various fluences. Average cluster radii ($<R> \approx 0.86$ -1.2 nm $\approx$ 16.3-22.7 a.u.), calculated using Eqn. (1), corresponding to $l = 0, 2$ are found to be nearly the same as expected and the mean values are inscribed in the Fig. 2 for the corresponding fluences of two different annealing conditions. Small sizes and narrow size distribution of the clusters grown in the present study may be attributed to the low ion current density adopted for the growth of the samples [8,9]. To determine the shape of the clusters, we performed Raman study in the *VH* and *VV* modes typically for the sample irradiated at a fluence of $1 \times 10^{20}$ m$^{-2}$ and annealed at 1473K (Fig.3a). Depolarization ratio ($I_{VH} / I_{VV}$) corresponding to $l = 2$ mode is observed to be ~0.70, which is close to 0.75 predicted for the spherical clusters with blue (488 nm in the present study) excitation [12]. Typical cluster radius of 1.25 (0.05) nm for 1473K annealed sample grown at a fluence of $1 \times 10^{21}$ m$^{-2}$ is independently determined by glancing incidence x-ray diffraction (GIXRD; STOE Diffractometer) analysis (Fig. 3b) using Scherrer's formula. The value matches well with 1.2 (0.01) nm measured from the low frequency Raman study (Fig. 2b) for the same sample.

    The increase in the cluster size with increasing fluence and annealing temperature can be understood as follows. Implanted Au is supersaturated in the c-$Al_2O_3$ matrix due to its low solubility. Fluctuations in the concentration nucleate gold atoms and these nuclei can grow directly

from the supersaturation. With increasing fluence, agglomeration to bigger clusters is likely [13]. However, sizeable growth of the cluster is likely with increasing annealing temperature and this may be due the fact that nuclei, which have reached a critical size, grow at the expense of smaller nuclei and known as the coarsening or ripening stage [13]. In the coarsening regime, nucleation of new particles is negligible. Thus, for any annealing temperature the cluster size is expected to increase with increasing fluence and for a particular fluence the cluster size should be larger with higher annealing temperature (Fig. 2).

For these spherical clusters in the smaller size range (inscribed in Fig. 2), a redshift is observed with decreasing cluster size (refer to experimental data for two different annealing temperatures in Fig. 4). Though the redshift effect is predicted to be quenched in case of free Au clusters, the large matrix-induced charge screening leads to a much larger electron spillout in case of embedded cluster [5]. Plasmon peak values, in our study, fall in between the values calculated using TDLDA method with $r = 0$ and 1 a.u. in the specific size limit. The redshift trend with decreasing cluster size is clearly depicted for $1>r\geq 0$ as the trend is reversed for $r>1$ with increased screening effect of $s$ and $d$ electrons [5]. Normally 'spillout' effect is realized in very small clusters where electron density reduces from its bulk value. Charge density for the clusters upto 170 atoms showed deviation from its bulk value in LDA calculations applied to the spherical jellium-background model (SJBM) [14], which later on improved up to 440 atoms in the TDLDA calculation [5]. In our study, with number of atoms, $N \sim 427$ (<440) for largest cluster of $\langle R \rangle \sim 1.2$ nm and down to 157 (inscribed in Fig. 2) for the smallest cluster of $\langle R \rangle \sim 0.86$ nm (~ 16.3 a.u.) definitely support our claim that the observed redshift of SPR peak with decreasing size is due to electron 'spillout' effect in the small Au clusters fully embedded in c-$Al_2O_3$ matrix. Unlike co-ablated growth of Au in alumina [4], the porosity effect leading to the blueshift SPR peak with decreasing cluster size might be negligible in our study as the samples were grown in a well crystalline $Al_2O_3$ matrix. Carrier escape by interband tunneling is also observed in semiconductor heterojunction nanostructure where electron dynamics of localized Wannier-Stark states [15], is extensively studied [16] in presence of electric field. However, dephasing of electron leading to a dynamic localization of electron in metallic nanoclusters may not be observed, as the coherence lifetime of electron (limited by scattering centers e.g., point defects and surface states in the nanocluster) will be much smaller than the period of oscillatory motion of electron for all reasonable value of applied electric field.

In conclusion, embedded Au nanoclusters of radii ~ 0.86-1.2 nm were grown by direct ion implantation in c-$Al_2O_3$ and subsequent air annealing treatments. Plasmon resonance spectra were recorded in the optical region of ~ 561-579 nm. A redshift of the plasmon peak with decreasing cluster size was observed. The redshift is correlated to the electron 'spillout' effect for the fully embedded Au clusters with number of atoms in the range of ~157-427.

The authors like to thank V. K. Indirapriyadarshini, and P. Ramamurthy of National Centre for Ultrafast Processes, University of Madras, India for their co-operation in UV-Vis studies. We also thank S. Kalavathi, MSD, IGCAR for the GIXRD study. We thank B. Viswanathan, MSD, IGCAR for his constant encouragement in pursuing this work.

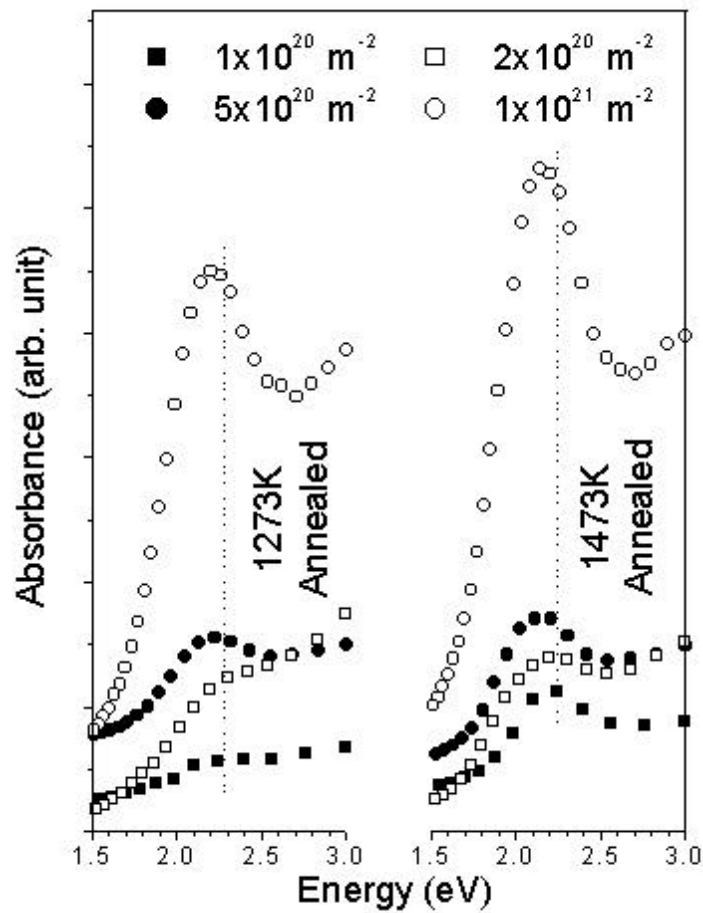

Fig. 1. SPR peaks for embedded Au nanoclusters in c-Al$_2$O$_3$ annealed at 1273K and 1473K showing redshift with decreasing fluence. Dashed vertical line is a guide to eye for the observed redshift of the SPR peak with decreasing fluence.

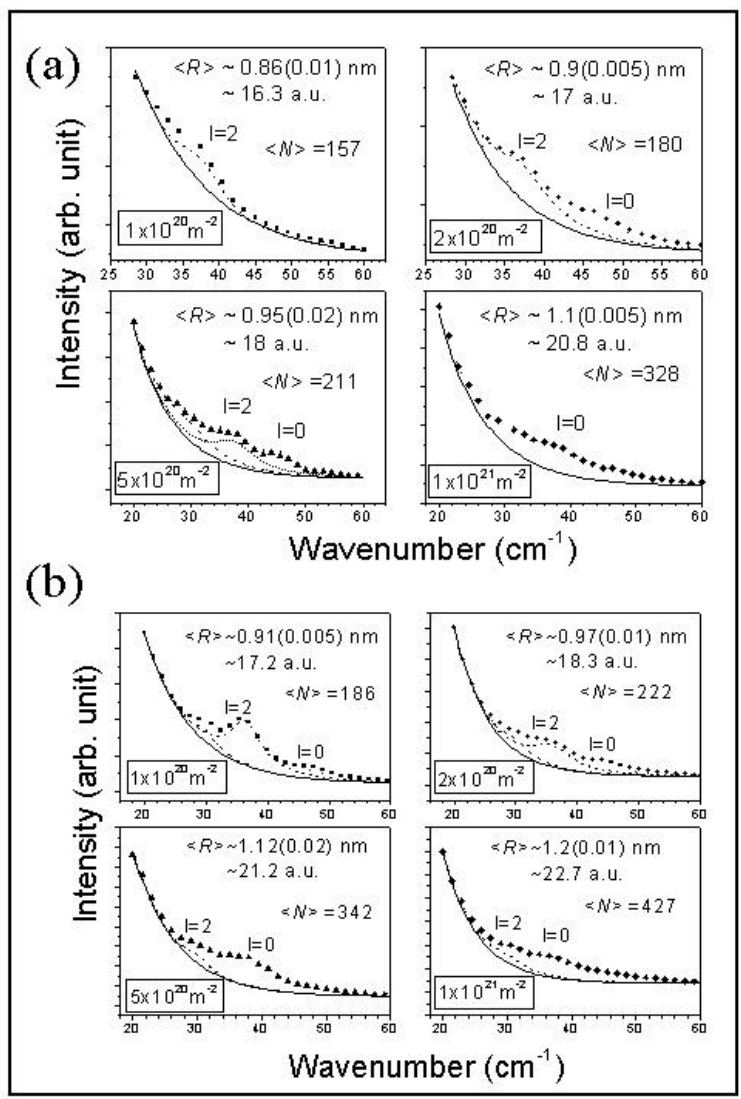

Fig. 2. Low-frequency Raman spectra for embedded Au nanoclusters in c-$Al_2O_3$ implanted at various fluences and annealed at (a) 1273K and (b) 1473K. Average cluster sizes (radii, *R*) determined from the eigenfrequencies corresponding to l = 0 and 2 (refer text) are inscribed for the corresponding fluences [Background (continuous lines); l = 0 (dotted lines) and l = 2 (dashed and dashed-dot lines)]. The radii are also presented in atomic unit (a.u.~ 0.0529 nm) and corresponding number of atoms, *N* are also calculated.

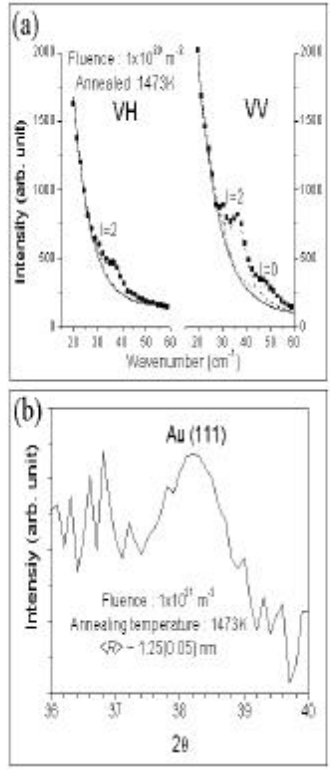

Fig. 3. (a) Low-frequency Raman spectra in *VV* and *VH* modes for the sample irradiated at a fluence of $1\times10^{20}$ m$^{-2}$ and annealed at 1473K [Background (continuous lines); $l = 0$ (dotted lines) and $l = 2$ (dashed and dashed-dot lines)]. (b) Typical GIXRD study of 1473K annealed sample grown at a fluence of $1\times10^{21}$ m$^{-2}$ showing Au (111) peak.

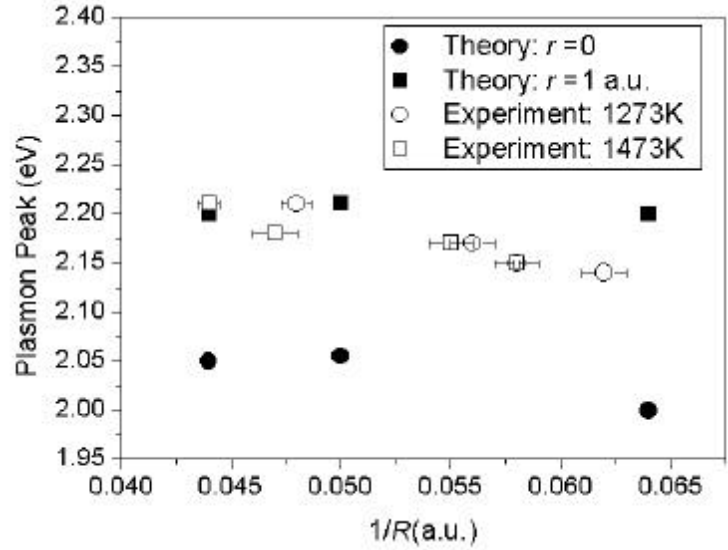

Fig. 4. Size dependence of the plasmon resonance peak for the samples annealed at 1273K and 1473K (unfilled symbols) and estimated values from TDLDA calculations (filled symbols) [extracted from Fig. 6(b) of Ref. 5] corresponding to $r = 0$ and 1 a.u. (refer text) are presented. Experimental values show a redshift with decreasing cluster size.